\documentclass[preprint]{aastex6}

\usepackage{bm}
\usepackage{lineno}
\nolinenumbers

\begin{document}
 
\title{{\bf 1-bit raw voltage recording system for dedicated observations of transients at low radio frequencies}}

\author{Kshitij S. Bane\altaffilmark{1},
Indrajit V. Barve\altaffilmark{2}, 
G. V. S. Gireesh\altaffilmark{2},
C. Kathiravan\altaffilmark{1}, and
R. Ramesh\altaffilmark{1}}
\altaffiltext{1}{Indian Institute of Astrophysics, Koramangala 2nd Block, Bangalore 560034, Karnataka, India}
\altaffiltext{2}{Radio Astronomy Field Station, Indian Institute of Astrophysics, Gauribidanur 561210, Karnataka, India}

\begin{abstract}
Recently we had reported commissioning of a prototype for pulsar observations at low radio frequencies ($<$100\,MHz)
using log-periodic dipole antennas (LPDAs) in the Gauribidanur Radio Observatory ($\rm {\approx}77^{\circ}E\,14^{\circ}N$) near Bangalore in 
India\footnote{\url{https://www.iiap.res.in/?q=centers/radio}}. The aforementioned system (GAuribidanur Pulsar System, GAPS) is currently being augmented to directly digitize the radio frequency signals from the individual antennas in the array. Our initial results using 1-bit raw voltage recording system indicates that such a back-end receiver offers distinct advantages like, (i) simultaneous observations of any set of desired directions in the sky with multiple offline beams and smaller data rate/volume, (ii) archival of the observed data with minimal resources for re-analysis in the future, either in the same or different set of directions in the sky.

\end{abstract}

\keywords{Radio observations: transients; Radio observations: pulsars;
Radio observations: instrumentation; Radio observations: low frequencies}

\section{Introduction} \label{sec:intro}
Latest advancements in commercial technology help to 
directly digitize the radio frequency (RF) signals in radio astronomy observations and record them as sampled voltages. Traditionally, a receiver architecture includes one or several mixer stages performing
down-conversion from RF to an intermediate or baseband frequency, due to limited analog-to-digital (ADC) sampling rate. However, in pace with technology development, increased ADC performance
allows for higher sampling rates. The higher the sampling rate, the higher the frequency that can be sampled for a correct read, thereby allowing sampling at either the observation frequency itself or a higher intermediate frequency. Accordingly, at a sufficiently high sample rate there would be no need for down-conversion. Eliminating the need of a mixer stage is advantageous, since mixers in practice raise system challenges, such as introducing frequency offsets. In addition to minimizing the number of analog components in the receiver system and thereby the associated amplitude/phase variations, the direct digitization offers more flexibility particularly in the efforts to search for radio transients at low frequencies where the individual dipole antennas have wider instantaneous sky coverage. For example, in the multi-beam setup where beams are formed on-board digital hardware like a FPGA, the number of beams is limited by the resources available on the FPGA. Also, once the beams are formed in certain directions, the observations are limited to the corresponding directions. On the contrary, if raw voltages from each antenna are recorded, any number of beams can be formed offline to probe any desired set of directions in the sky within the field-of-view (FoV) of the individual antenna. Therefore, it is advantageous to have a voltage recording capability. A high sampling rate and high resolution ADC accordingly approaches an accurate read of an incoming signal, seemingly preferable. However, the energy consumption at high sampling frequencies increases with
the number of quantization levels\footnote{\url{https://github.com/bmurmann/ADC-survey}}. Furthermore, the resulting data volumes and subsequent data processing/management will be enormous in such a system. 
One solution to these is to record fewer number of bits from analog-to-digital converters connected to each antenna element.  Data rates and volumes can be significantly reduced if 1-bit (sign bit) digital receivers are employed. 1-bit digital correlators have been used in radio astronomy \citep[see e.g.][]{Weinreb1963,Uday1990,Nakajima1994,Ebenezer2001,
Ramesh2008,Zakharenko2016}. The present work describes a 1-bit raw voltage capture system that we have implemented in  GAPS to demonstrate how such a system could be useful for observations of pulsars and other transients at low radio frequencies.

\section{The antenna and analog receiver system} \label{sec:ana}

\begin{figure}[t!]
\centerline{\includegraphics[width=15cm]{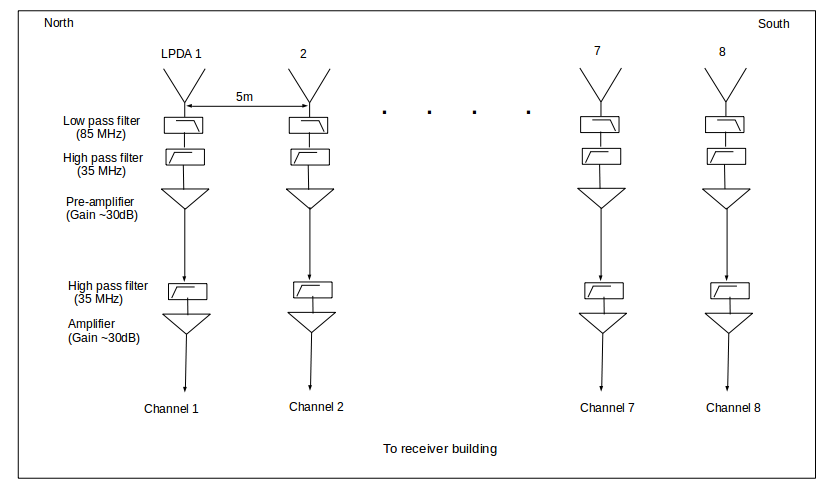}}
\caption{GAPS array configuration used in the present work.}
\label{fig:figure1}
\end{figure}

GAPS is a recently commissioned prototype using LPDAs for observations of the transients in the frequency range ${\approx}$35\,-\,85\,MHz. The antenna array has 16 LPDAs arranged on a north-south baseline with spacing of 5\,m between the adjacent antennas \citep{Bane2022}. For the present work, we used only 8 LPDAs (see Figure \ref{fig:figure1}). The half-power beam width (HPBW) of each LPDA in the GAPS are ${\approx}80^{\circ}$ in the E-plane where the arms of the LPDA are present, and 
${\approx}110^{\circ}$ in the orthogonal H-plane. All the LPDAs have been mounted with their H-plane in the east-west direction and E-plane in the north-south direction. The effective collecting area and gain of each LPDA is ${\approx}0.4{\lambda}^{2}$ (where ${\lambda}$ is the wavelength corresponding to the observing frequency) and ${\approx}$6.5\,dBi (with respect to an isotropic radiator), respectively. The characteristic impedance of the LPDA is ${\approx}$50\,${\Omega}$. The Voltage Standing Wave Ratio (VSWR) is $<$\,2 over the frequency range ${\approx}$40\,-\,440\,MHz. Since the LPDAs are arranged on a north-south baseline, the combined response pattern of the array (8 LPDAs) after coherent beamforming is ${\approx}110^{\circ}$ in the east-west (hour angle/right ascenssion) direction and ${\approx}6^{\circ}$ in the north-south (declination) direction for observations near the zenith at a typical frequency like 60 MHz. It is a fan-beam with the above resolution
in declination. The width of the east-west response pattern is nearly independent of frequency. Being very wide, it helps to observe a radio source continuously for ${\approx}$7\,h. This is useful since mechanical steering is difficult for the dipole antennas used at low frequencies
\citep[see e.g.][]{Ramesh1999b,Ramesh2012b}. LMR-200 coaxial
cable\footnote{\url{https://timesmicrowave.com/wp-content/uploads/2022/06/lmr-200-datasheet.pdf}} connected to the feedpoint near the top of the LPDA transmits the RF signal incident on the LPDA to the input of a low-pass filter with 3 dB cut-off at ${\approx}$85\,MHz followed by a high-pass filter with 3\,dB cut-off at ${\approx}$35\,MHz, and then a wideband amplifier with an uniform gain 
${\approx}$30\,dB in the frequency range 35\,-\,85\,MHz. The two filters and the amplifier are kept near the base of the LPDA to minimize the length of the LMR-200 cable and hence the transmission loss. 
The high- and low-pass filters help to attenuate the unwanted signal at frequencies ${\lesssim}$30\,MHz and 
${\gtrsim}$85\,MHz. Both the filters have steep reduction in gain beyond their cut-off frequencies. For e.g. the amplitude of the input signal is reduced by 
${\approx}$30\,dB at 25\,MHz and 95\,MHz in the aforementioned high- and low-pass filter, respectively. Due to this, amplification of any radio frequency interference (RFI) present outside the frequency range of our observation (i.e. 35\,-\,85\,MHz) is very minimal. The passband is relatively free of RFI \citep[see e.g.][]{Bane2022}. 
The output corresponding to each LPDA are then transmitted to a central cabin near the center of the array where they are again high-pass filtered and amplified. Note that the 2nd high-pass filter in the signal chain further suppresses the RFI at frequencies ${\lesssim}$30\,MHz. The RF signals are then transferred to the receiver building (located 
${\approx}$300\,m away) independently via eight separate, low-loss LMR-400 coaxial cables\footnote{\url{https://timesmicrowave.com/wp-content/uploads/2022/06/lmr-400-datasheet.pdf}} of identical RF characteristics. The lengths of all the cables are the same and they were also phase equalized. The cables are buried 
${\approx}$1\,m below the ground level to minimize possible diurnal variations in their characteristics. 

The array was initially equipped with an analog beamformer network \citep{Bane2022} where phase/delay cables controlled by diode switches were used to observe a particular direction of the sky at any given epoch \citep[see e.g.][]{Landecker1984,Ramesh1998}. Different lengths of phase/delay cables must be used in the latter to observe different regions of the sky. This introduces systematic gain variations. Furthermore, simultaneous observations are practically difficult. Once the beam is formed in a particular direction, observations will be restricted to that direction. These are common to any other similar analog hardware used elsewhere. Note that we used only 8 LPDAs in the GAPS for the present work since our aim is to understand the digital beamformer described in the present work for future use with the Gauribidanur RAdioheliograPH (GRAPH) which has 8 LPDAs per group \citep{Ramesh2011a,Ramesh2014}, and use GRAPH for dedicated observations of non-solar transients particularly during the local night time. Further, the analog-to-digital convertors available with us can also handle 8 inputs only. Note that the GRAPH is a T-shaped radio interferometer array consisting of 512 LPDAs. The individual arms of the array are oriented along the east-west and south directions. The effective collecting area ($\rm A_{e}$) of the array is 
$\rm {\approx}6400m^{2}$ at a typical frequency like 60\,MHz. The sky coverage is ${\approx}$8800\, square degrees, independent of frequency. Both these numbers are reasonably larger, an important requisite for observations of non-solar transients. For comparison, the fully completed NenuFAR in France is expected to have $\rm A_{e}{\approx}20000m^{2}$ and sky coverage ${\approx}100$ square degrees, at 60\,MHz\footnote{\url{https://nenufar.obs-nancay.fr/en/astronomer/}}. The corresponding numbers for LOFAR which is currently operational are $\rm A_{e}{\approx}9600m^{2}$ and sky coverage ${\approx}19600$ square degrees, at 60\,MHz \citep{Haarlem2013}. The aforementioned characteristics of GRAPH, along with the prospects of dedicated and continuous observing period of 
${\approx}$14\,h every day exclusively for non-solar transients suggests that GRAPH could be an useful instrument.

\section{The digital receiver system} \label{sec:rec}

\begin{figure}[t!]
\centerline{\includegraphics[width=15cm]{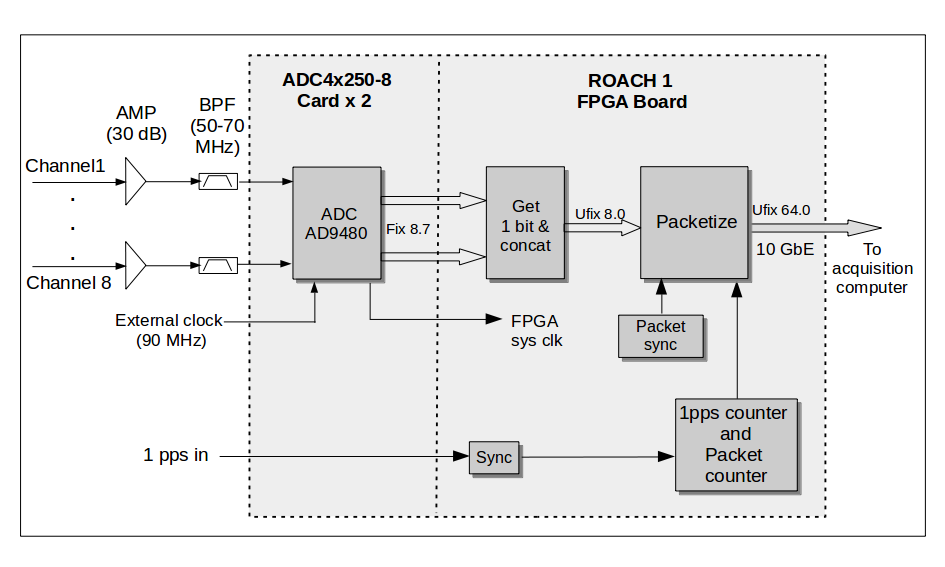}}
\caption{1-bit raw voltage recording system used in the GAPS.}
\label{fig:figure2}
\end{figure}

Figure \ref{fig:figure2} shows the schematic of the 1-bit raw voltage recording receiver. The digital receiver is implemented on Reconfigurable Open Architecture Computing Hardware (ROACH) from the Collaboration for Astronomy Signal Processing and Electronics Research (CASPER). The ROACH board hardware has the Xilinx Virtex-5 FPGA\footnote{\url{https://casper.astro.berkeley.edu/wiki/ROACH}}. The RF signals from the each of the eight antennas pass through independent amplifier and bandpass filter of 50\,-\,70 MHz before being fed to ADCs. We have used two Quad-ADC cards, each containing four AD9480 ICs\footnote{\url{https://www.analog.com/en/products/AD9480.html}}. The signals are sampled at 90\,MHz. A single clock is used to sample all the antennas.
The sampled band lies in the second Nyquist zone of 90\,MHz. The ADC converts the input voltage to a 8-bit Fixed point number (Fix 8.7) between -1 and +1. The sign bit from each of these numbers is separated. 
So we have 1-bit from each antenna channel. Therefore, the ADC output is an unsigned 8-bit number (UFix8.0) representing the input voltage from each of the 8 channels. 
Such 1024x8 8-bit numbers are packetized together and sent to the recording computer over 10 Gigabit Ethernet (10GbE). Note that 10GbE allows transmission of 64-bit numbers\footnote{\url{https://casper-toolflow.readthedocs.io/projects/tutorials/en/latest/tutorials/roach/tut{\_}ten{\_}gbe.html}}. So, one packet has 1024 64-bit numbers. This results in a data rate of 
${\approx}$0.72 Gbits/s. This is much smaller compared to the the data rate of ${\approx}$5.6 Gbits/s which would have resulted if 8-bits were recorded from each antenna channel. To avoid packet losses while recording, the data is captured using n2disk program\footnote{\url{https://www.ntop.org/products/traffic-recording-replay/n2disk/}} on a 32 GB RAM computer. Each packet also contains a 
user datagram protocol (UDP) header(58 bytes) and a custom header (80 bytes) that contain the observation details, 1 pulse per second (PPS) count, and a packet count. 
The 1 PPS signal is derived from a GPS clock\footnote{\url{https://timing.trimble.com/wp-content/uploads/thunderbolt-e-gps-disciplined-clock-datasheet.pdf}} and given to the FPGA via the ADC card to generate the 1 PPS count. The packet-count is a unique packet number assigned to each packet. It is used to check for any packet losses. The recorded file contains a time series of 1-bit voltages from each antenna channel. Some of the characteristics are listed in Table 1.

\begin{table}[!t]
\centering
\caption{Characteristics of GAPS}
\label{tab:table1}
\begin{tabular}{ll}    
\\ 
\hline
Number of channels & 8 \\
Sampling rate & 90\,MSPS \\
Sampled bits & 1 \\
Observation band & 45\,-\,90 MHz (50\,-\,70 MHz for the present work) \\
Beamforming & offline \\
FFT size & Variable (8192 for the present work) \\
Frequency resolution & 10.98\,kHz \\
Post-detection time resolution & Variable (min 91.02\,${\mu}$s for one spectrum) \\
Declination coverage & $-26.4^{\circ}$\,S to $+53.6^{\circ}$\,N \\
{\bf 1} & {\bf 2} \\
\hline
\end{tabular}
\end{table}

\section{Offline beam formation} \label{sec:off}

\begin{figure}[t!]
\centerline{\includegraphics[width=15cm]{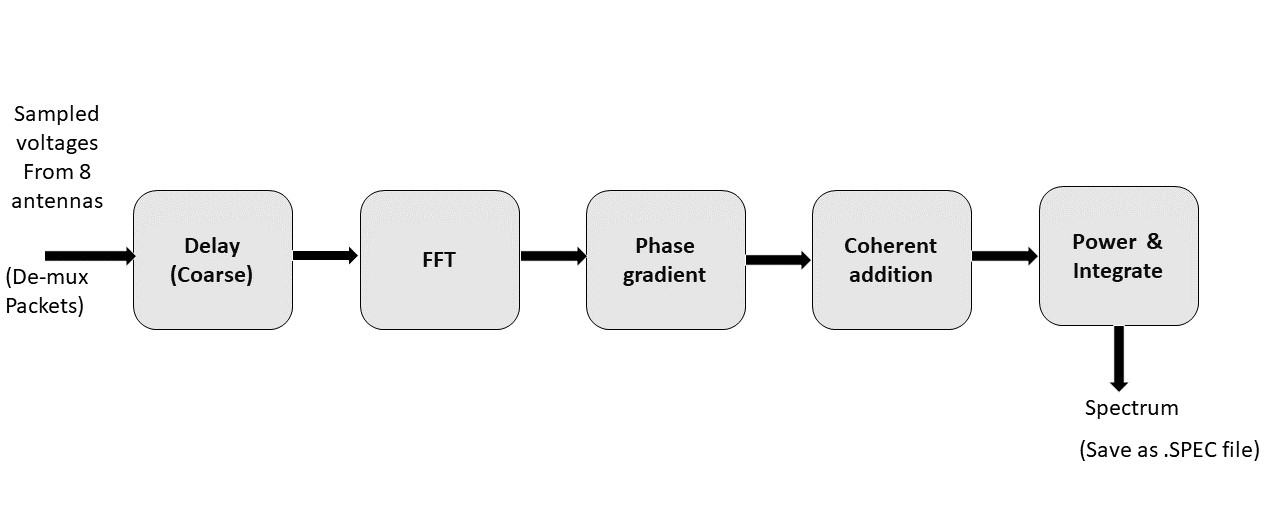}}
\caption{Offline beamforming pipeline used in the GAPS.}
\label{fig:figure3}
\end{figure}

Figure \ref{fig:figure3} shows the offline beamforming pipeline. First, the packets are unscrambled to get voltage time series of the individual antenna channels. Delays are applied to these time series according to the desired declination where the beam needs to be formed. Note that the array is oriented in the north-south direction, so the beam forming is performed along the declination. Based on the declination, the geometric delays are calculated using   
the formula $t_{j}$=$d_{j}sin{\theta}/c$, where $t_{j}$ is the delay required for $j^{th}$ antenna at a distance of $d_{j}$ from the reference antenna for the source at angle ${\theta}$, and $c$ is the velocity of light. The sampling interval ($t_{s}$) is 1/90\,MHz = 11.11\,ns. So, the time series are delayed by the closest integer multiple of 11.11\,ns (i.e. $mt_{s}$, where $m$ is an integer). 
The uncompensated fractional delays (${\Delta}t_{j}$=$t_{j}$\,-\,$mt_{s}$) can be compensated by applying a phase gradient in the spectral domain, i.e.
${\phi}_{j,k}$=$2{\pi}F_{k}{\Delta}t_{j}$, where $F_{k}$ is $k^{th}$ frequency bin. The delayed voltage streams are Fourier transformed using FFT. Subsequenty, the phase compensation is performed by multiplying the complex Fourier output by $e^{-i{\phi_{j}}}$. An 8192-point FFT is performed, resulting in 4096 positive frequency bins with a resolution of 10.986\,kHz. The number of FFT points and hence the resolution can be changed depending on the requirement since the beamforming is performed offline. After the phase compensation, the individual spectra from each antenna are added together, modulo-squared, and integrated according to the desired temporal resolution. Multiple beams pointing at different declinations can be formed by changing the delay and phase values.  A `.SPEC' file is stored for each beam containing the corresponding power spectrum series. These are further processed using the standard PulsaR Exploration and Search TOolkit \citep[PRESTO,][]{Ransom2011}. 

\section{Observations} \label{sec:anares}

\begin{figure}[t!]
\centerline{\includegraphics[width=20cm]{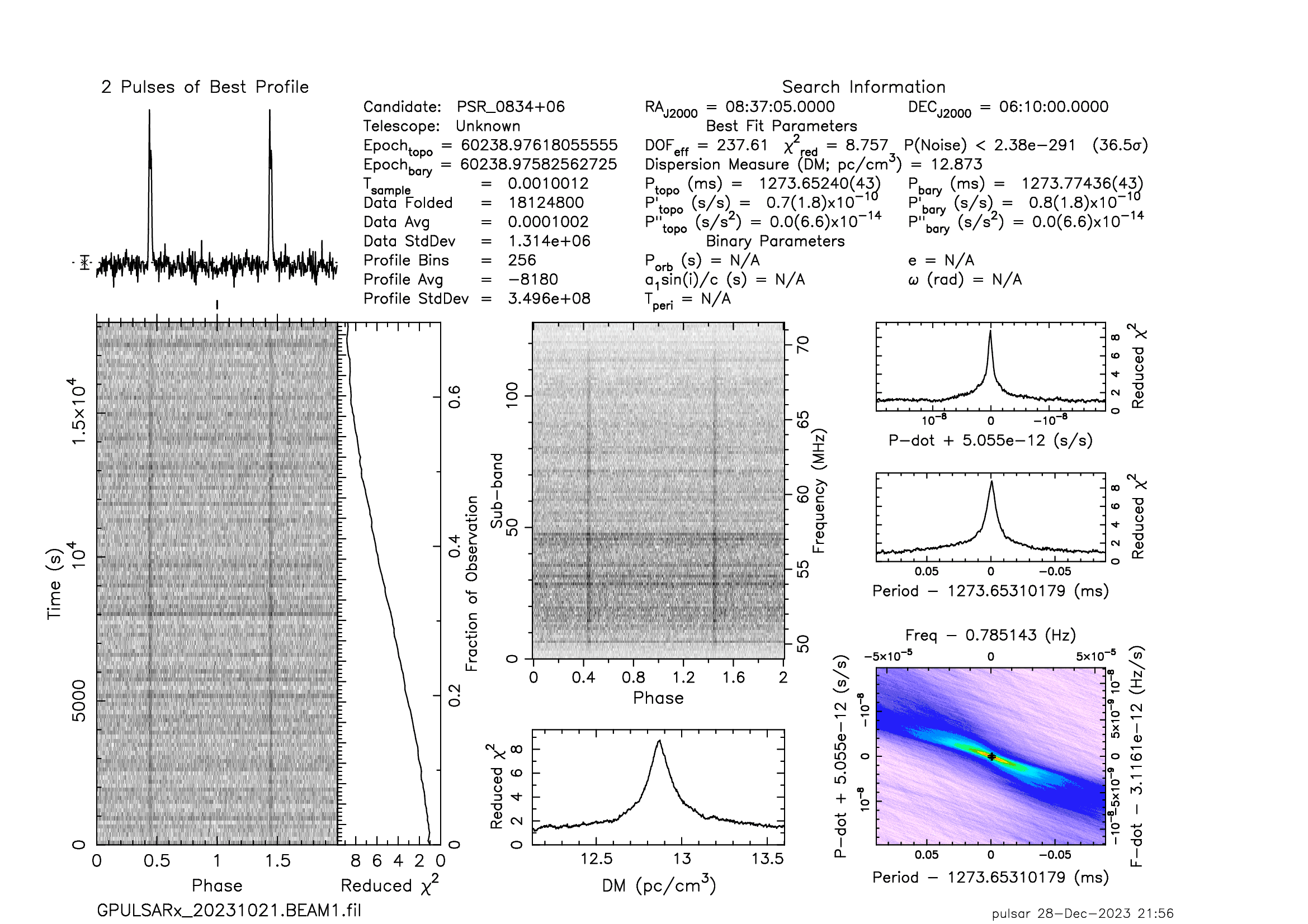}}
\caption{Observations of the pulsar B0834+06 (J0837+0610) on 2023 October 21 with GAPS in the frequency range of 50 to 70 MHz for a period of ${\approx}$4\,h. The data were analyzed using PRESTO.}
\label{fig:figure4}
\end{figure}

\begin{figure}[t!]
\centerline{\includegraphics[width=20cm]{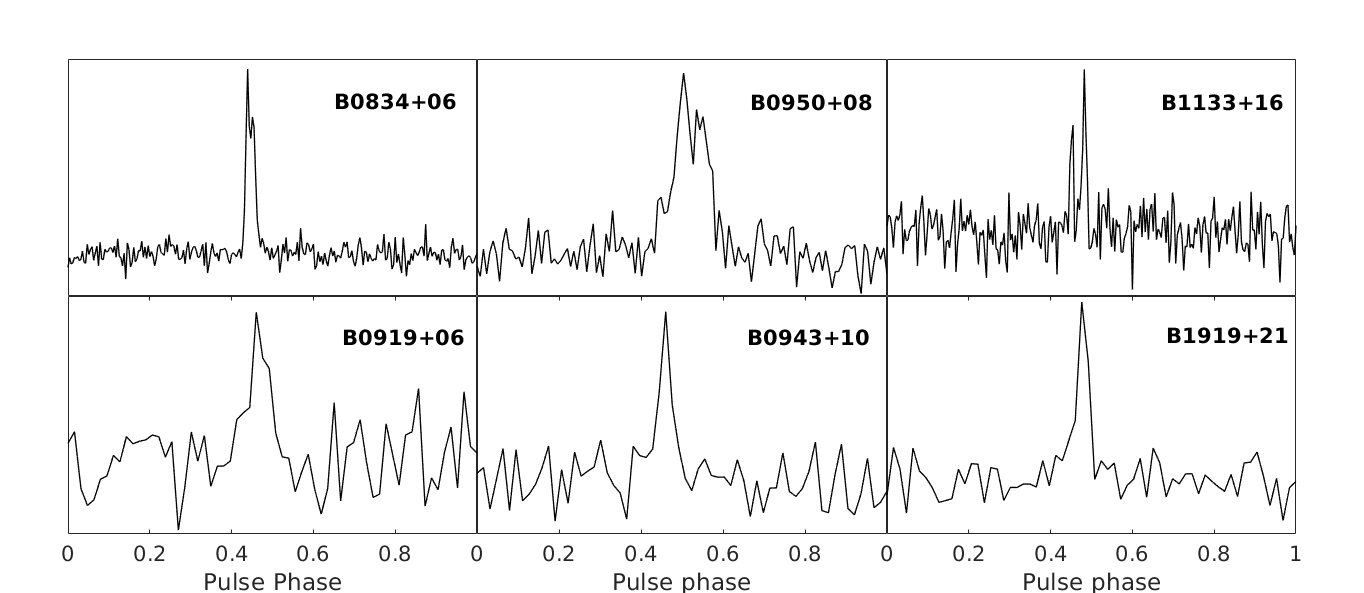}}
\caption{Average profiles of the pulsars observed with GAPS using 1-bit raw voltage recording system and offline beamformation.}
\label{fig:figure5}
\end{figure}

\begin{figure}[t!]
\centerline{\includegraphics[width=20cm]{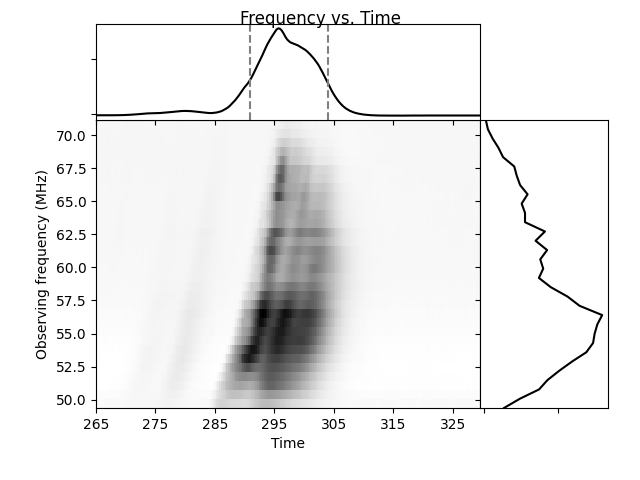}}
\caption{
De-dispersed dynamic spectrum (50 to 70 MHz) corresponding to the type III solar radio burst observed with the GAPS on 2023 August 10. The DM used was $\rm {\approx}13.2\,pc\,cm^{-3}$. The time in the x-axis starts from ${\approx}$11:02:30\,LST, and the
interval is 10\,s. The upper panel show the spectrally averaged time profile of the burst. The right
panel shows the temporally averaged spectral profile of the burst. The dotted lines in the upper
panel indicate the main phase of the burst.}
\label{fig:figure6}
\end{figure}

The initial run of the new system was carried out on 2023 August 10 for 
${\approx}$5\,h, starting from ${\approx}$08:49\,LST. Three beams pointing towards declinations $6.17^{\circ}$\,N, $7.92^{\circ}$\,N, and $15.85^{\circ}$\,N were formed offline from the raw voltage data recorded. The power spectrum series  from these three beams were stored in three separate .SPEC files with time resolutions of 4\,ms, 2\,ms and 1\,ms, respectively. The files were converted to SIGPROC `filterbank' (.fil) format\footnote{\url{http://sigproc.sourceforge.net/sigproc.pdf}}. As first step, RFI excision was performed using 
{\it rficlean} \citep{Maan2021} to mitigate strong periodic RFI. Subsequently we used {\it rfifind} from PRESTO, which searches for time and frequency domain RFI and creates mask files. This is followed by {\it prepfold} which is used to fold and dedisperse the data and detect pulsars. Three pulsars B0834+06 (J0837+0610), B0950+08 (J0953+0755) and B1133+16 (J1136+1551) were detected in the above three beams, respectively (see Figures \ref{fig:figure4} \& \ref{fig:figure5}). 
A solar event \citep[type III burst,][]{McLean1985,Ramesh2001b} was also recorded during the observation at ${\approx}$11:03\,LST.
The declination of Sun was ${\approx}15.7^{\circ}$\,N. A separate beam was formed pointing towards the above declination with time resolution of ${\approx}$10\,ms. 
We carried out analysis with single pulse search pipeline developed using PRESTO routines. The maximum SNR was obtained for 
dispersion measured (DM)\,${\approx}$\,13.2\,$\rm pc\,cm^{-3}$. The de-dispersed dynamic spectra of the solar burst is shown in Figure \ref{fig:figure6}. The burst was noticed in the observations carried out with the solar radio spectrographs in the Gauribidanur observatory also \citep{Benz2009,Kishore2014}. 

\begin{figure}[t!]
\centerline{\includegraphics[width=20cm]{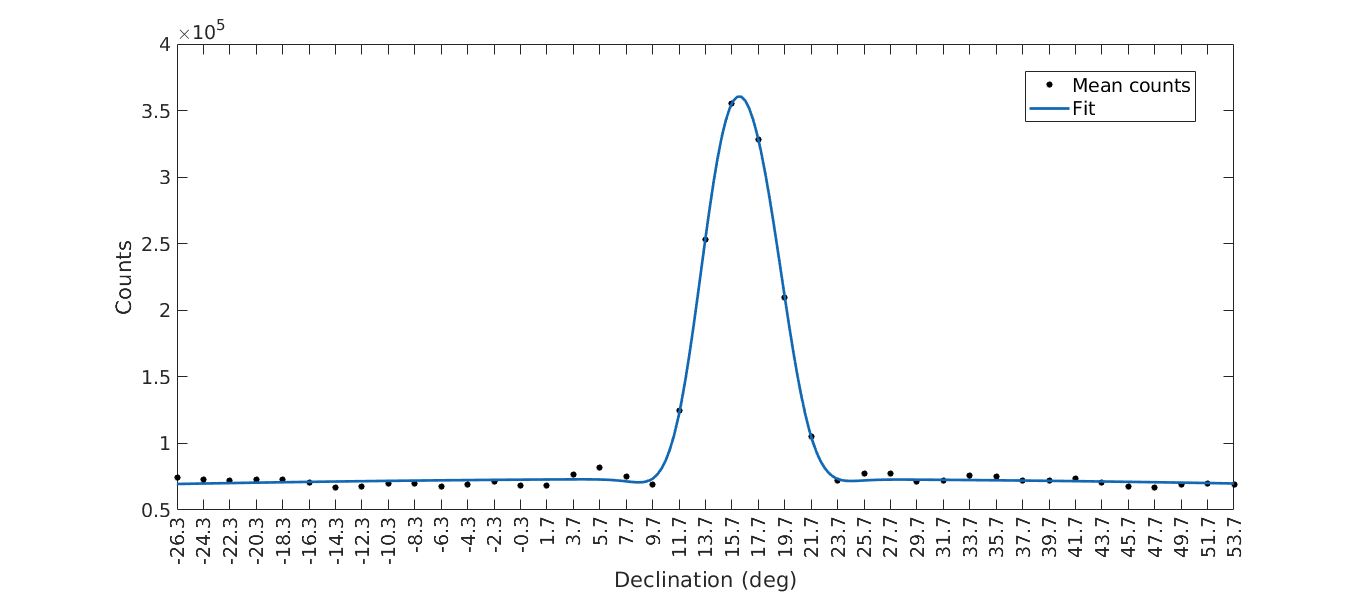}}
\caption{The declination beam of GAPS at 60\,MHz obtained by using multiple beams formed during the solar observations in Figure \ref{fig:figure6}.}
\label{fig:figure7}
\end{figure}

To verify the performance of GAPS, we formed 41 beams covering the declination range from $-26.3^{\circ}$\,S to $+53.7^{\circ}$\,N in steps of $2^{\circ}$ for the duration of the solar burst in Figure \ref{fig:figure6}. Mean count during the solar burst at 60\,MHz in each of the 41 beams were obtained. The count is maximum for the beam closest to the Sun’s declination. A Gaussian fit to the data points indicates HPBW ${\approx}6^{\circ}$ (see Figure \ref{fig:figure7}), which matches with the predicted HPBW of the GAPS antenna array in the present case (see Section 2). On  2023 October 18, a 4\,h observation run starting at ${\approx}$17:27\,LST was carried out. A beam was formed pointing towards declination ${\approx}21.88^{\circ}$\,N with 4\,ms time resolution. Pulsar B1919+21 (J1921+2153) was detected in this data (see Figure \ref{fig:figure5}). A subsequent observation for a period of ${\approx}$7.2\,h was performed on 2023 October 21 starting at ${\approx}$06:36\,LST.  Pulsars B0834+06 (J0837+0610), B0919+06 (J0922+0638), B0950+08 (J0953+0755), B0943+10 (J0946+0951) and B1133+16 (J1136+1551) were detected using ${\approx}$6\,h data around the transit of each pulsar (see Figure \ref{fig:figure5}). Four beams were formed pointing towards declinations $6.17^{\circ}$\,N, 
$7.92^{\circ}$\,N, $9.86^{\circ}$\,N and $15.85^{\circ}$\,N with 1\,ms resolution. Pulsars B0834+06 and B0919+06 were detected in the same beam. The estimated parameters of the different pulsars mentioned above are listed in Table 2. The corresponding numbers reported previously by \cite{Bondonneau2020} from online, multi-bit digitization observations elsewhere in nearly the same frequency range are mentioned within the brackets. The period \& DM for each pulsar in the two cases agree well. The pulsar average profiles in Figure \ref{fig:figure5} also closely match with the average profiles at low frequencies reported in the literature \citep{Malov2010,Hassall2012,Zakharenko2013,Stovall2015,Pilia2016,Bilous2020, Bondonneau2020}. 
Any minor differences are likely due to noisy detections in the present case. For example in the case of \cite{Bondonneau2020}, the bandwidth and $\rm A_{e}$ used were higher than the present observations by factors of 
${\approx}$2.75 \& ${\approx}$10, respectively. Furthermore, 12-bit ADC was used in their observations. Its quantization efficiency is 
${\approx}$(1/0.64) times higher than that of the 1-bit ADC used in the present observations \citep{Thompson2001}. The duration of observation for the respective pulsars in the two cases are also different. If we take into consideration all these factors, then the expected SNR for the pulsar B0834+06 in GAPS observations as compared to \cite{Bondonneau2020} is,
${\frac{309*0.64}{10{\sqrt{2.75/6}}}}{\approx}30$.
Compared to this, the observed SNR with GAPS is ${\approx}$36.
Note that the duty cycle mentioned in Table 1 corresponds to the effective pulse width in pulse profiles at 50\% of the highest peak intensity (i.e. w50).

\section{Conclusions} \label{sum}

We have presented a 1-bit raw voltage recording system for observations of pulsars at low radio frequencies. Other than reduction in the senstivity which could be partly compensated by the use of higher sampling rate \citep{VanVleck1966,Burns1969,Thompson2001},
the 1-bit quantization does not have an effect on the parameters of a pulsar like period, DM, etc. (see Table 2).
Most of the modern 
low-frequency radio telescopes have multi-bit digital receivers which provide higher dynamic range \citep{Taylor2012,Haarlem2013,Prabu2015,Reddy2017,Zarka2020,Girish2023}. However, in a relatively minimal RFI location like Gauribidanur \citep{Monstein2007,Kishore2015,Hariharan2016b,Bane2024}, a lower dynamic range digital receiver can be used \citep[see e.g.][]{Zakharenko2016}. Our results indicate that in such circumstances the sensitivity to observe faint events within a short time interval depends largely on the effective collecting area and the observing bandwidth.
It is well known that absolute flux measurments can be difficult in 1-bit receivers
\citep{VanVleck1966,Uday1990,Ramkumar1994,Ramesh2006b,Stein2019}. But in a wide-field multi-beam system where the primary requirement is to cover maximum area of the sky with high temporal and spectral resolution, such binary receiver systems can be beneficial.
Overall, the 1-bit receiver offers flexibility, allowing multiple simultaneous beams in different directions, different integration times and frequency channel widths, and different filtering algorithms to be used. 
That wouldn't make your purposes less compelling but it would certainly advocate for this kind of recording mostly being used for bright events.
Further, the archival data can be stored in raw form with fewer resources and reprocessed in the future as per requirements.
Reports indicate that 1-bit raw voltage recording had been used at higher frequencies also \citep{Teng2015}. With the currently available technology, sampling rates upto ${\approx}$40\,GHz can be achieved.

\begin{table}[!t]
\centering
\caption{Characteristics of the pulsars observed with GAPS using 1-bit raw voltage recording system
(Numbers within the brackets are from \cite{Bondonneau2020})}
\label{tab:table2}
\begin{tabular}{llllll}     
\noalign{\smallskip}\hline\noalign{\smallskip}
Pulsar & Period & DM & Duty cycle & SNR & Observing \\
& (s) & ($\rm pc/cm^{3}$) & (\%) & (folded) & duration \\
& & & & & (h) \\
\hline
B0834+06 & 1.274${\pm}0.000043$ & 12.873${\pm}0.05$ & 6.4${\pm}0.8$ & 36.5 & 6 \\
& (1.274) & (12.864) & (6.9) & (309) & (1) \\
B0919+06 & 0.431${\pm}0.000016$ & 27.330${\pm}0.02$ & 14.4${\pm}0.8$ & 6.0 & 6 \\
& (0.431) & (27.296) & (15.4) & (144) & (3) \\
B0943+10 & 1.098${\pm}0.000069$ & 15.348${\pm}0.02$ & 12.9${\pm}0.8$ & 6.3 & 6 \\
& (1.098) & (15.329) & (15.2) & (148) & (2.5) \\
B0950+08 & 0.253${\pm}0.000050$ & 2.965${\pm}0.05$ & 15.3${\pm}0.8$ & 24.6 & 6 \\
& (0.253) & (2.971) & (14.6) & (140) & (1) \\
B1133+16 & 1.188${\pm}0.000023$ & 4.848${\pm}0.03$ & 12.7${\pm}0.8$ & 8.4 & 6 \\
& (1.188) & (4.846) & (18.7) & (261) & (2) \\
B1919+21 & 1.337${\pm}0.000017$ & 12.420${\pm}0.02$ & 7.8${\pm}0.8$ & 12.5 & 4 \\
& (1.337) & (12.437) & (8.4) & (180) & (1) \\
\hline
\end{tabular}
\end{table}
 
We are grateful to Gauribidanur Observatory team for their help in the observations and upkeep of the facilities. A. A. Deshpande is thanked for his encouragement and suggestions. We acknowledge the referee for his/her kind comments, which helped to present the results more clearly.



\end{document}